\documentstyle[preprint,prl,aps,epsfig]{revtex}
\tightenlines

\begin{document}

%%%%%%%%%%%%%%%%%%%%%%%%%%%%%%%%begin text%%%%%%%%%%%%%%%%%%%%%%%%%%%%%%%%%%%%
\preprint{\vbox{\hbox{\underline{\bf revised manuscript}}
                \hbox{BNL-68328}
                \hbox{PRINCETON/HEP/2001-1}
                \hbox{TRI--PP--01--09}
                \hbox{KEK Preprint 2001-26}}
          }

%=======================================================================
% begin my own macros
\newcommand{\kpigamma}{\mbox{$K^+ \! \rightarrow \! \pi^+ \gamma$\ }}
\newcommand{\kmutwo}  {\mbox{$K^+ \! \rightarrow \! \mu^+ \nu$\ }}
\newcommand{\kmutwog} {\mbox{$K^+ \! \rightarrow \! \mu^+ \nu \gamma$\ }}
\newcommand{\kmuthre} {\mbox{$K^+ \! \rightarrow \! \mu^+ \pi^0 \nu$\ }}
\newcommand{\kethre}  {\mbox{$K^+ \! \rightarrow \! e^+   \pi^0 \nu$\ }}
\newcommand{\pnn}   {\mbox{$K^+ \! \rightarrow \! \pi^+ \nu \overline{\nu}$\ }}
\newcommand{\kpipi}   {\mbox{$K^+ \! \rightarrow \! \pi^+ \pi^0$\ }}
\newcommand{\kpitwo}  {\mbox{$K_{\pi 2}$\ }}
% end my own macros
%=======================================================================

%==============================================================================

\title{Search for the rare decay \kpigamma}

%%%%%%%%%%%%%%%%%%%%%%begin author list%%%%%%%%%%%%%%%%%%%%%%%%%%%

\author{
S.~Adler$^1$, M.~Aoki$^6$\cite{masa}, M.~Ardebili$^5$, M.S.~Atiya$^1$,
A.O.~Bazarko$^5$,
P.C.~Bergbusch$^{6,8}$, E.W.~Blackmore$^6$, D.A.~Bryman$^{6,8}$,
I-H.~Chiang$^1$, M.R.~Convery$^5$\cite{markc}, M.V.~Diwan$^1$, J.S.~Frank$^1$,
J.S.~Haggerty$^1$, T.~Inagaki$^3$, M.M.~Ito$^5$\cite{marki}, V.~Jain$^1$, 
S.~Kabe$^3$, M.~Kazumori$^3$\cite{UT},
S.H.~Kettell$^1$, P.~Kitching$^7$, M.~Kobayashi$^3$,
T.K.~Komatsubara$^3$, A.~Konaka$^6$, Y.~Kuno$^3$\cite{kuno}, M.~Kuriki$^3$,
T.F.~Kycia$^1$\cite{ted}, K.K.~Li$^1$, L.S.~Littenberg$^1$,
J.A.~Macdonald$^6$, R.A.~McPherson$^5$\cite{robm},
P.D.~Meyers$^5$, J.~Mildenberger$^6$, M.~Miyajima$^2$,
N.~Muramatsu$^3$\cite{UT}\cite{mura}, T.~Nakano$^4$,
C.~Ng$^1$\cite{susb}, J.~Nishide$^2$,
T.~Numao$^6$, A.~Otomo$^3$\cite{UT}, J.-M.~Poutissou$^6$, 
R.~Poutissou$^6$, G.~Redlinger$^6$\cite{george}, T.~Sasaki$^4$,
T.~Sato$^3$, T.~Shinkawa$^3$\cite{shink}, F.C.~Shoemaker$^5$, 
R.~Soluk$^7$, J.R.~Stone$^5$, R.C.~Strand$^1$,
S.~Sugimoto$^3$, Y.~Tamagawa$^2$,
C.~Witzig$^1$, and Y.~Yoshimura$^3$
\\ (E787 Collaboration) }

\address{ 
$^1$Brookhaven National Laboratory, Upton, New York 11973\\
$^2$ Department of Applied Physics, Fukui University,
 3-9-1 Bunkyo, Fukui, Fukui 910-8507, Japan\\  
$^3$High Energy Accelerator Research Organization (KEK), 
Oho, Tsukuba, Ibaraki 305-0801, Japan \\ 
$^4$RCNP, Osaka University, 10-1 Mihogaoka, Ibaraki, Osaka 567-0047, 
Japan \\ 
$^5$Joseph Henry Laboratories, Princeton University, Princeton, 
New Jersey 08544 \\
$^6$ TRIUMF, 4004 Wesbrook Mall, Vancouver, British Columbia,
Canada, V6T 2A3\\
$^7$ Centre for Subatomic Research, University of Alberta, Edmonton,
Alberta, Canada, T6G 2N5\\
$^8$ Department of Physics and Astronomy, University of British Columbia, 
Vancouver, BC, Canada, V6T 1Z1
}

%%%%%%%%%%%%%%%%%%%%%%end author list%%%%%%%%%%%%%%%%%%%%%%%%%%%
%\date{}

\maketitle
\begin{abstract}

We have performed
a search for the angular-momentum forbidden decay \kpigamma
with the E787 detector at BNL.
No events were observed in the $\pi^+$ kinematic region around 227 MeV/$c$.
An upper limit on the branching ratio for the decay is 
determined to be 
$3.6\times 10^{-7}$ at 90\% confidence level. 
\end{abstract}
\pacs{PACS numbers: 11.30.Cp, 13.25.Es}
%%%
%%% 11.30.Cp Lorentz and Poincare invariance
%%%	in 11.30.-j Symmetry and conservation laws 
%%% 13.25.Es Decays of K mesons
%%%	in 13.25.-k Hadronic decays of mesons
%%%		in 13. Specific reactions and phenomenology
%%%

\draft
%-----------------------------------------------------------------------------
\pagebreak

\section{Introduction}

 We report on a search for the decay \kpigamma.  
 This is 
 a spin $0 \to 0$ transition with a real photon  
 and is thus forbidden by angular momentum conservation~\cite{textbook}. 
 This decay is also forbidden 
 on gauge invariance grounds~\cite{gauge}.
 Historically, the absence of \kpigamma relative to the \kpipi decay (\kpitwo)
 led Dalitz
 to determine the kaon spin 
 to be zero rather than two or greater~\cite{Dalitz}. 
 In 1969,
 a model of strange particles and weak decays 
 predicted this decay at a branching ratio 
 of $2\times 10^{-4}$~\cite{Selleri}; 
 the model was later ruled out 
 by an experimental upper limit of $4\times 10^{-6}$
 at 90\% confidence level (C.L.)~\cite{Klems}.
  
 No theory of physics beyond the Standard Model,
 if it is based on point-particle quantum field theory,
 allows \kpigamma decay.
 Current interest in this decay
 stems from speculation that
 an experimental signature of exotic physics, such as
 a vacuum expectation value of a new vector field~\cite{KaneNP}, 
 non-local Superstring effects~\cite{KanePT}, or
 departures from Lorentz invariance~\cite{ColemanGlashow},
 could appear in this decay mode. 
 No specific theoretical prediction or bound on the branching ratio
 has yet been given.

 In previous experiments~\cite{Klems,Asano3}
 no candidate events were detected.
 The most recent upper limit of $1.4 \times 10^{-6}$ 
 at 90\% C.L.~\cite{Asano3}
 was established in 1982.

\begin{figure}
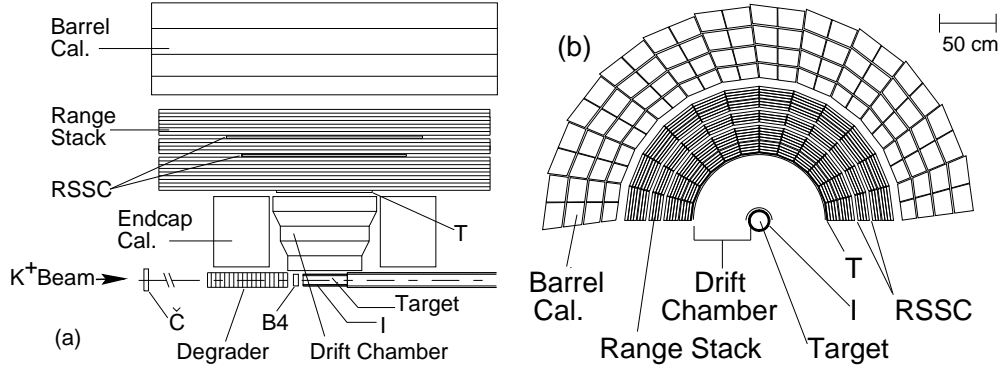

\hfill
\centerline{
\hbox{\psfig{figure=e787kpgrev_fig1a_e787_side.epsi,width=6.5cm}}
\hbox{\psfig{figure=e787kpgrev_fig1b_e787_end.epsi,width=6.5cm}}
}
\hfill
\caption{Schematic side-view (a) and end-view (b)
         showing the upper half of the E787 detector.
         \v{C}: \v{C}erenkov counter; 
         B4: energy-loss counter; 
         I and T: trigger scintillation counters (I-counters and T-counters);
         RSSC: range-stack straw-tube tracking chambers.} 
\label{fig:E787sch}
\end{figure}

 The new search reported here used 
 the E787 detector~\cite{E787det0} (see Fig.~\ref{fig:E787sch})
 at the Alternating Gradient Synchrotron (AGS) 
 of Brookhaven National Laboratory (BNL).
 E787 is a 
 rare kaon-decay experiment studying 
 \pnn ~\cite{E787pnn} and related decays~\cite{E787rel},
 using kaon decays at rest. 
 \kpigamma
 is a two-body decay with 
 a 227-MeV/$c$ $\pi^+$ track
 and a 227-MeV photon emitted directly opposite 
 to it.
 It is assumed
 that energy-momentum is conserved 
 or 
 that its violation due to exotic physics 
 is tiny and undetectable in this decay mode.

 The main background source 
 is \kpitwo decay
 with a branching ratio of 0.2116~\cite{PDB}
 and a $\pi^+$ momentum of 205 MeV/$c$.
 The higher energy photon ($>$ 125MeV) 
 from the $\pi^0\to\gamma\gamma$ decay in \kpitwo
 tends to be emitted 
 opposite the $\pi^+$ track.
 \kpitwo background events can survive 
 if the $\pi^+$ momentum is mis-measured to be too large 
 and at the same time  
 the lower energy photon from the $\pi^0$
 (or the electron-positron pair from the $\pi^0\to\gamma e^+ e^-$ decay)
 is undetected. 
 However, 
 the redundant kinematic measurements and 
 efficient photon detection
 available in the E787 detector 
 are suited to 
 suppress the \kpitwo and other backgrounds
 to well below the sensitivity for the signal.

\section{Detector}

The AGS delivered kaons of about 700 MeV/$c$ to the experiment 
at a rate of $(4-7) \times 10^{6}$ per 1.6-s spill.
The kaon beam line~\cite{E787lesb3} incorporated two stages of
electrostatic particle separation, 
which reduced the pion contamination 
to 25\% at the entrance of the detector.
Kaons were detected and identified by a \v{C}erenkov counter,
multi-wire proportional chambers and an energy-loss counter.
After being slowed by a BeO degrader, approximately 25\% of the incident kaons
came to rest in an active stopping target
located at the center of the detector.
The 12-cm diameter target, which 
consisted of 0.5-cm square plastic-scintillating fibers,  
provided initial tracking of the 
stopping kaon and 
its decay products.

Particles emanating from 
kaon decays at rest in the target were measured
in a solenoidal spectrometer with a 
1.0-T field along 
the beam axis.
The charged decay products passed through
a layer of plastic scintillation counters 
surrounding the target (I-counters) and 
a cylindrical drift chamber~\cite{E787utc}, 
and lost energy in an array
of plastic scintillation counters called the Range Stack (RS).
The drift chamber
provided tracking information for momentum determination 
from 12 layers of 
axial anode-wire cells and six layers of thin 
spiral-strip cathode foils
with a total mass of $3\times 10^{-2}$ radiation lengths.
The RS provided a measurement of range and kinetic energy 
of the $\pi^+$ track which came to rest in it.
The radial region between 45.1 cm and 89.6 cm for the RS
was segmented into 24 azimuthal sectors 
and 21 radial layers totaling one radiation length.
The RS counters in the first layer (T-counters),  
which defined the solid angle 
acceptance 
for the $\pi^+$ track in the RS,
were 0.635-cm thick and 52-cm long;
the subsequent RS counters in the second layer 
(RS2-counters) and beyond
were 1.905-cm thick and 182-cm long.
The RS counter in the sector and layer where the $\pi^+$ track came to rest
was called the ``stopping counter''. 
All the RS counters were read out by phototubes 
on the upstream and downstream ends.
The output pulse shapes
were recorded by 500-MHz sampling transient digitizers (TDs)~\cite{E787td},
each of which was based on 
two interleaved 250-MHz 8-bit flash 
analog-to-digital converters (ADCs).
In addition to providing precise time and energy information
for reconstructing the $\pi^+$ track, 
the TDs enabled us to observe the $\pi^+\to\mu^+\nu$ decay at rest 
in the RS stopping counter. 
Two layers of straw-tube tracking chambers were embedded
in between the 10th and 11th RS layers and 
between the 14th and 15th RS layers, respectively.

A hermetic calorimeter system, 
 designed primarily to detect photons from \kpitwo and other decays
 in the \pnn  study,  
surrounded the central region.
The cylindrical barrel (BL) calorimeter 
covering about two thirds of the solid angle, 
and located immediately outside the RS, 
was used to search for the photon from the \kpigamma decay.
It consisted of 
alternating layers of lead (0.1-cm thick)
and plastic scintillator (0.5-cm thick) sheets and 
was segmented azimuthally into 48 sectors whose 
boundaries were tilted so that 
the inter-sector gaps 
did not project back to the target. 
In each sector, 
four radial groups of 
16, 18, 20 and 21 lead-scintillator layers, respectively
with increasing radius,
formed BL modules totaling
14.3 radiation lengths.
About 29\% of the shower energy
was deposited in the scintillators.
The BL modules, which were 190-cm in length along the beam axis, 
were read out by phototubes on the upstream and downstream ends,
and the outputs were recorded by 
time-to-digital converters (TDCs) and ADCs.
The two endcap calorimeters~\cite{E787endcap} and
additional calorimeters for filling minor openings along the beam direction,
as well as any active parts of the detector
not hit by the $\pi^+$ track,
were used for detecting extra particles including photons.
The two endcap calorimeters consisted of 143 undoped-CsI crystals,
which were read out by fine-mesh phototubes in the 1.0-T field
into 500-MHz TDs based on charged coupled devices~\cite{E787ccd}.

\section{Trigger}

The signature of \kpigamma in the experiment
was a two-body decay of a kaon at rest with 
a 227-MeV/$c$ $\pi^+$ track in the RS 
and a 227-MeV photon emitted directly opposite 
to it
and observed as a single cluster in the BL calorimeter.
The \kpigamma trigger
required a kaon decay at rest,
followed by a $\pi^+$ track which came to rest in the RS and
a shower cluster in the BL, 
and no extra particles in the BL, endcap or RS counters. 

A kaon was identified in the trigger
by a coincidence of hits 
from the \v{C}erenkov counter, energy-loss counter and target.
The timing of the outgoing pion (via the I-counters) 
was required to be at least 1.5 nsec later than
timing of the incoming kaon (via the \v{C}erenkov counter).
This online delayed-coincidence requirement 
guaranteed that the kaon actually decayed at rest, and
removed contributions 
from beam pions that were scattered into the detector and 
from kaons that decayed in flight after the \v{C}erenkov counter.
A single charged track was required 
to have a coincidence of the hits 
from the I-counters and 
from the T-counter and 
RS2-counter 
in the same RS sector.
The track was also required to penetrate 
to at least the sixth RS layer
in the sector with the coincidence
of T-counter and RS2-counter hits 
or in either of the next two clockwise RS sectors,
 in order to select positively charged particles.
Tracks reaching the outer three RS layers 
(from the 19th to the 21st layers) were rejected,  
in order to suppress the muons from \kmutwo and \kmutwog decays.
A $\pi^+\to\mu^+\nu$ decay at rest in the RS stopping counter
was identified online,
based on the pulse shape information from the TDs on the RS,
by an algorithm
demanding either a clear double pulse or an
RS pulse with an area larger than expected from its height
due to the presence of the 4 MeV $\mu^+$ from the $\pi^+$ decay.
The events that failed the algorithm were rejected,   
in order to further reduce muon tracks as well as electron tracks.
The number of shower clusters in the BL calorimeter
with energy above 10 MeV
was counted online
and was restricted to be one.
An event was rejected
if the energy observed in the endcap calorimeter 
was more than 20 MeV
or the energy in the RS sectors outside the region 
of the $\pi^+$ track
was more than 10 MeV.

The trigger was pre-scaled by $2450$ 
for taking data simultaneously with the trigger for \pnn.
This resulted in a total exposure of kaons entering the target 
available for the \kpigamma search to be $6.7\times 10^{8}$. 
A total of $1.6\times 10^{6}$ events
met the trigger requirements. Most of them were due to \kpitwo .

\section{Offline Analysis}

\subsection{Event reconstruction}

In the offline analysis, 
the momentum ($P$),
the range (equivalent cm of plastic scintillator, $R$)
and the kinetic energy ($E$)
of the $\pi^+$ track
were reconstructed
with the target, drift chamber and RS information.
The momentum was determined 
by correcting the measured momentum in the drift chamber
for the energy loss suffered by the $\pi^+$ track
with the observed track length in the target.
The range was calculated from the track lengths in the target
and in the RS.
The kinetic energy was determined by adding up the energy deposits 
of the $\pi^+$ track
in the scintillators of the target and the RS, 
taking account of the losses 
in inactive materials such as wrapping and chamber components.
 The RS energy calibration was done for the upstream and downstream 
end separately, 
 using the energy loss in each counter of muons from
 \kmutwo decay.
The kinematic resolutions 
(rms) were 
$\Delta P =$ 2.6 MeV/$c$, 
$\Delta R =$ 1.28 cm and 
$\Delta E =$ 3.8 MeV. 

The timing, 
energy ($E_{\gamma}$) and 
direction of the photon
were determined from reconstruction of the hits in the BL calorimeter 
and the kaon decay vertex position in the target.
The $\pi^+$ track in the RS defined the event time reference, 
and photons from \kpitwo decay
were used to calibrate
the BL hit time.
The energy calibration of the BL hits was performed 
 using the energy depositions from cosmic rays, 
 and was verified by reconstructing the energy of the $\pi^0$
 in \kpitwo decay. 
The $z$ position (along the beam axis) of the BL hits was measured
with TDC and ADC information from phototubes
on both ends of the BL modules,
and was used in conjunction with target information to determine the
polar angle of the photon. 
The energy resolution of the BL calorimeter was 
$\Delta E_{\gamma}/E_{\gamma}=6\%/\sqrt{E_{\gamma}}$  ($E_{\gamma}$ in GeV).
The resolutions of azimuthal and polar opening angles 
between the photon and the $\pi^+$ track,
$\phi_{\pi^+\gamma}$ and $\theta_{\pi^+\gamma}$,
were determined to be 
$\Delta \phi_{\pi^+\gamma} =2.2^{\circ}$ and
$\Delta \theta_{\pi^+\gamma} =3.9^{\circ}$, respectively.

\subsection{Primary cuts}

The following selection criteria (``cuts''),
called the primary cuts in the analysis, 
were imposed
on the events that were successfully reconstructed.

A stopping-layer cut accepted
only events whose $\pi^+$ track came to rest in the RS layers
beyond the inner RS tracking chamber
(from the 11th to the 18th layers) for analysis. 
This cut was imposed to ensure that the range measurement
included positions determined by the RS tracking chambers.

Taking into consideration 
the limited energy resolution and segmentation of the BL calorimeter,
the cuts imposed on the energy and direction of the photon in the BL were 
relatively loose:
${E_{\gamma}\geq 120}$ MeV, 
${|\phi_{\pi^+\gamma}|\geq 165^{\circ}}$ and
${\theta_{\pi^+\gamma}\geq 165^{\circ}}$.
A coincidence cut ($\pm$ 2 ns) 
between the times of the photon and the $\pi^+$ track
was imposed.
This cut greatly 
reduced the occurrence of photon candidates due to accidental hits. 
Events in which the photon cluster in the BL
showed associated activity in the neighboring RS sectors were rejected
by a ``RS preshower'' cut, so that 
only those events whose
total photon shower energy was deposited in the BL calorimeter
were accepted. 

To remove events triggered by kaon decays in flight
or by multiple beam particles into the detector, 
an offline delayed-coincidence cut 
requiring $>$ 2 ns 
between the pion time and
the kaon time measured in the target,  
and cuts 
on the timing and energy of the hits recorded 
in the \v{C}erenkov counter, proportional chambers, 
energy-loss counter and target
were imposed. 

A double-pulse fit to the pulse shape recorded in the RS stopping counter
was made offline to 
identify pions with
$\pi^+\to\mu^+\nu$ decay at rest; 
this cut removed contamination
from \kmutwo , \kmutwog , \kmuthre and \kethre decays, 
as well as \kpitwo decays 
whose $\pi^+$ 
decayed in flight before it came to rest in the RS.

\subsection{$\pi\gamma$ sample and $\pi\pi^0$ sample}

The following two data sets were made
from the events that survived the primary cuts.

The set of events with 
${218\leq P \leq 234}$ MeV/$c$
was identified as the
``$\pi\gamma$ sample''.
The \kpigamma signal region
\footnote{
 This signal region is the same as that for
 the search for $K^+\to\pi^+ X^{0}$ decay in \cite{E787pnn}, 
 where $X^{0}$ is 
 a neutral weakly interacting massless particle~\cite{familon} and
 the $\pi^+$ momentum, 227 MeV/$c$, 
 is identical to the momentum for \kpigamma .}
was further specified as the subset of the $\pi\gamma$ sample 
with 
${35.5\leq R \leq 40.0}$ cm and
${120\leq E \leq 135}$ MeV.
These cuts
ensured that 
the $\pi^+$ momentum, range and kinetic energy were consistent with those
of the $\pi^+$ track from the \kpigamma decay: 
$227$ MeV/$c$, $38.5$ cm and $127$ MeV, respectively.

The set of events that were collected by the same \kpigamma trigger
and whose $\pi^+$ momentum and range were rather consistent 
with \kpitwo decay
 (${197.5\leq P \leq 212.5}$ MeV/$c$ and 
  ${27.0\leq R \leq 35.0}$ cm)
was identified as the ``$\pi\pi^0$ sample''. 
This sample 
was rich in \kpitwo background, and
was used for evaluating the rejections of cuts 
in the background studies.

\section{Background Studies}

After the primary cuts were imposed, 
the remaining background events were mostly from \kpitwo decay. 
This background was due to the disappearance of the softer of the two
photons from  $\pi^0$, 
either through inefficiency due to 
very narrow gaps between counters, inactive material, etc.
(``photon detection inefficiency'') 
or through overlap with the charged track (``overlapping photon'').
Both types
were studied
from the data by establishing 
two independent sets of offline cuts for each type;  
one set consisted of 
the cuts on the $\pi^+$ momentum, range and kinetic energy,
and the other set consisted of ``photon veto cuts''
(for photon detection inefficiency)
or ``$dE/dx$ cuts''
(for overlapping photon), as explained in the next subsections. 
In these studies
we selected events in the $\pi\pi^0$ sample or
inverted at least one of these cuts on the events in the $\pi\gamma$ sample,
in order to enhance the background collected by the \kpigamma trigger
as well as to prevent candidate events from being examined 
before the background studies were completed.
In order to avoid contamination from other background sources, 
all the offline cuts except for those being established
were imposed on the data
\footnote{
 For example, in the study of the photon detection inefficiency,
 $dE/dx$ cuts were imposed 
 on both the $\pi\pi^0$ sample and the $\pi\gamma$ sample in advance.
 }.

\subsection{Photon detection inefficiency}

In the offline analysis, photon shower activity was identified
in the various subsystems, including the BL calorimeter and the RS,
as hits in the counters 
in coincidence with the $\pi^+$ track within a few ns 
and with energy above a low threshold (typically $\sim$1 MeV).
Events with 
extra activity not associated with the $\pi^+$ and 
the candidate signal photon 
were rejected by the offline photon veto cut of each subsystem.

\begin{figure}
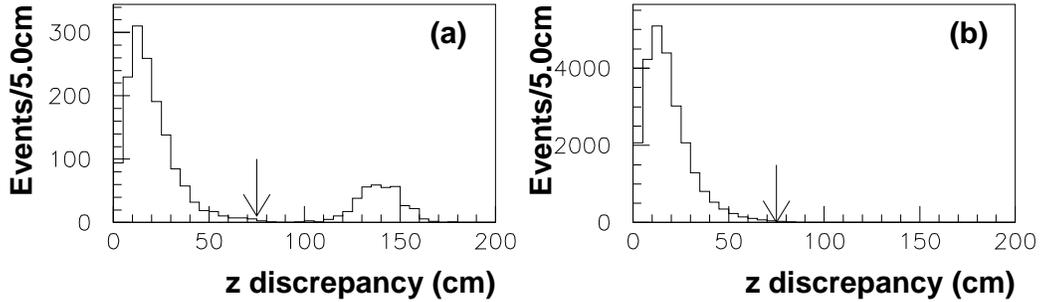

\hfill
\centerline{
\hbox{\psfig{figure=e787kpgrev_fig2rev_a.epsi,height=4.0cm}}
\hbox{\psfig{figure=e787kpgrev_fig2rev_b.epsi,height=4.0cm}}
}
\hfill
\caption{
         (a) $z$ discrepancy distribution
         of the BL clusters 
         in the subset of the $\pi\pi^0$ sample
	 with all photon veto cuts imposed.
         (b) $z$ discrepancy distribution of the BL clusters 
          when the other photon from the $\pi^0$ decay in \kpitwo was detected
          outside of the cluster.
          Events above the arrow in each distribution
          corresponded to multiple photon hits and were rejected by
          the cut at $75$ cm.} 
\label{fig:bvc}
\end{figure}

If two photons from 
a $\pi^0$ hit the same or adjacent BL modules, 
they form a single high-energy BL cluster opposite the $\pi^+$ track 
and can mimic \kpigamma decay.  In such cases,
due to the kinematics 
of \kpitwo and subsequent $\pi^0\to\gamma\gamma$ decays,
the two photons must hit the modules at different $z$ positions
along the beam axis.
The coincidence cut between the times of the photon and the $\pi^+$ track,
already imposed as a primary cut, 
reduced the number of events with two photons in the same BL module,
because the mean of the arrival times of the 
scintillation light at the upstream and downstream ends of the BL module
was significantly earlier than that of a single photon hit.
 Furthermore, 
 the BL hits in the cluster were examined and the maximum
        discrepancy among the $z$-position measurements obtained
	from TDC and ADC information was determined.
Fig.~\ref{fig:bvc}~(a) shows the $z$ discrepancy distribution
of the BL clusters in the subset of the $\pi\pi^0$ sample
that survived all photon veto cuts.
Comparing this to Fig.~\ref{fig:bvc}~(b), 
which shows the $z$ discrepancy distribution of the BL clusters 
when the other photon from the $\pi^0$ decay in \kpitwo was detected
outside of the cluster,
the events with a large discrepancy in Fig.~\ref{fig:bvc}~(a)
clearly represent the case when two photons hit adjacent BL modules.
A ``BL cluster'' cut was therefore employed
to reject an event
if the discrepancy was more than $75$ cm.

To evaluate the background rejection 
of the offline photon veto cuts including the BL cluster cut,  
the subset of the $\pi\pi^0$ sample whose $\pi^+$ kinetic energy
was also consistent with \kpitwo decay (${97\leq E \leq 117}$ MeV) was 
examined.
By imposing the photon veto cuts on this subset, 
the background rejection of these cuts was measured to be $13.8$ 
as shown in Fig.~\ref{fig:summary_kpi2}~(a).
Then, in the subset of the $\pi\gamma$ sample
that failed at least one of the offline photon veto cuts,
the kinematic distributions of \kpitwo background events
\footnote{
  The $\pi\gamma$ sample was made by choosing the events 
    that survived the $\pi^+$ momentum cut
    ${218\leq P \leq 234}$ MeV/$c$.}
was studied.
The $\pi^+$ range versus kinetic energy plot of these events
is shown in Fig.~\ref{fig:summary_kpi2}~(b); 
there are no events in the \kpigamma signal region, and 
287 events are consistent with \kpitwo decay in range and kinetic energy.
This indicates that there are no mis-measured tracks 
in or around the signal region, and
a further background reduction by $13.8$ is expected,
assuming the $\pi^+$ range and kinetic-energy cuts and the photon veto cuts 
are independent.

In estimating the background level of \kpitwo 
in a specific $\pi^+$ region in range and kinetic energy,
the number of events in the region in Fig.~\ref{fig:summary_kpi2}~(b)
was divided by the rejection (13.8) minus 1
 \footnote{
    The number of events should be divided 
     by the tagging efficiency of the inverted photon veto cuts $\epsilon$,
     and by the rejection of the photon veto cuts $R$.
    Since $\epsilon$ is equal to $1-\frac{1}{R}$,
     the estimate is equivalent to the number of events divided by $R$ minus 1.
}.
The number of events in the signal region in Fig.~\ref{fig:summary_kpi2}~(b)
was taken to be $<2.44$ events at 90\% C.L.~\cite{PDB,FeldmanCousins}
instead of zero. 
The background levels of \kpitwo in the signal region
and in the \kpitwo region were therefore estimated to be 
$< 0.19$ events at 90\% C.L. and $22.4\pm 1.3$ events, respectively.

Fig.~\ref{fig:summary_kpi2}~(b) indicates that
no correlation between the $\pi^+$ range and kinetic energy
is visible 
in the \kpitwo background events
at the current sensitivity of the search. 
If we further added the assumption that 
the $\pi^+$ range and kinetic energy measurements were indeed not correlated, 
the background level of \kpitwo in the signal region
was estimated to be $< 0.0007$ events at 90\% C.L.
\footnote{
  The number of events that survived the $\pi^+$ range cut, 
   $< 2.44$ events at 90\% C.L. instead of zero,
  was divided by the rejection of the $\pi^+$ kinetic-energy cut   
  (taken to be $287$ from the number of events 
   in Fig.~\ref{fig:summary_kpi2}~(b)) 
  minus $1$, and by the rejection of the photon veto cuts
  ($13.8$) minus $1$.
}

\subsection{Overlapping photon}

The above
background study could possibly be confounded by \kpitwo events in which
the shower of the lower energy photon from the $\pi^0$ overlapped 
some of the counters hit by the $\pi^+$ track.
The reconstructed kinetic energy of such tracks could be
incorrectly measured due to additional energy deposited in the scintillators
by the overlapping photon.
A  set of $dE/dx$ cuts, 
which checked the consistency between the measured energy and range 
in each of the RS counters,  
was therefore employed to reject this type of background.
Events with a RS counter in which the measured energy was 
larger than expected from the reconstructed range in that counter
were rejected by the $dE/dx$ cuts.

To evaluate the background rejection of the $dE/dx$ cuts, 
we selected events in the $\pi\pi^0$ sample whose $\pi^+$ kinetic energy
was larger than that from \kpitwo decay 
and was in the signal region for \kpigamma 
(${120\leq E \leq 135}$ MeV).
By imposing the $dE/dx$ cuts on this subset, 
the background rejection of these cuts was measured to be $36.3$ 
as shown in Fig.~\ref{fig:summary_overlap}~(a).
From the subset of the $\pi\gamma$ sample
that failed at least one of the $dE/dx$ cuts, 
the $\pi^+$ range versus kinetic energy plot
is shown in Fig.~\ref{fig:summary_overlap}~(b); again 
no mis-measured tracks are in or around the \kpigamma signal region. 
The background level of \kpitwo in the signal region
was estimated to be $< 0.07$ events at 90\% C.L.
\footnote{
  The number of events in the signal region 
  in Fig.~\ref{fig:summary_overlap}~(b), 
  $< 2.44$ events at 90\% C.L. instead of zero,
  was divided by the rejection of the $dE/dx$ cuts ($36.3$) minus $1$.} 
The estimate is limited by statistics.

\subsection{$K^+\to\pi^+\gamma\gamma$ decay}

  For $K^+\to\pi^+\gamma\gamma$ decay
  in the $\pi^+$ momentum region greater than 215 MeV/$c$, 
  a 90\% C.L. upper limit of $5.0\times 10^{-7}$ on the branching ratio
  has been established~\cite{pigg} (assuming a phase-space kinematic 
  distribution).
  Taking into account further suppression 
  by the trigger requirements, 
  offline photon veto cuts and $dE/dx$ cuts 
  to the $K^+\to\pi^+\gamma\gamma$ decay, 
  its contribution to the \kpigamma search 
  is negligible at the current sensitivity.

\section{Result}

\subsection{Events in the signal region}

\begin{figure}
\hfil
\centerline{
  \psfig{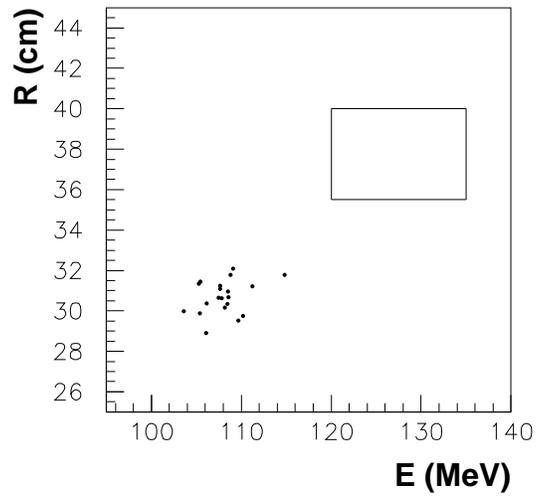}
}
\hfill
\caption{
  Range versus kinetic energy plot of the events 
  with all analysis cuts imposed. 
  The box indicates the signal region for \kpigamma .} 
\label{fig:finalplot}
\end{figure}

Fig.~\ref{fig:finalplot} shows the $\pi^+$ range versus kinetic energy plot
of the events that survived all analysis cuts. 
No events were observed in the signal region. 
The group of 
20 events around E = 108 MeV was due to the \kpitwo background
and was consistent with the $22.4\pm 1.3$ events
expected from the estimate
of the photon detection inefficiency discussed above.

\subsection{Sensitivity}

The single-event sensitivity for \kpigamma decay in this search
was obtained by normalizing 
to the number of \kpitwo events 
collected by the \kpigamma trigger.
For the \kpitwo events,  
the $\pi^+$ track in the RS and the higher energy photon 
in the BL calorimeter were reconstructed, 
and the offline analysis cuts of the \kpigamma search 
except for those sensitive to the photons from $\pi^0$
(the RS preshower cut, 
 photon veto cuts, $dE/dx$ cuts and target energy cuts) were imposed.
The number of events whose $\pi^+$ momentum, range and kinetic energy 
were consistent with 
\kpitwo decay was
$3.62\times 10^{5}$.

Acceptance factors 
were determined 
from the sample generated by Monte Carlo simulation
and from 
the data samples of \kpitwo decays
\footnote{
 The sample of \kpitwo decays for measuring acceptance factors 
 in this subsection
 was accumulated by a calibration trigger
 that removed the requirements
 on the shower and visible energy 
 in the BL, endcap and RS counters
 from the \kpigamma trigger.}, 
\kmutwo decays and scattered beam pions,
which were accumulated by calibration triggers
simultaneous to the collection of signal candidates.
Many systematic uncertainties in the measurement of the 
acceptance factors for \kpigamma and \kpitwo 
(e.g., the fraction of kaons entering the target that decayed at rest,
 $\sim 0.72$)
canceled in taking the ratio of acceptances of 
these decay modes.
The factors are summarized in Table~\ref{tab:acc_pgamma}.

The acceptance factors 
of the $\pi^+$ reconstruction cuts 
 (including the online requirements in the trigger) 
and the $\pi^+$ kinematic fiducial cuts 
 specifying the signal region and the \kpitwo region
 (including the stopping-layer cut)
were estimated primarily from Monte Carlo simulation.
The loss when the $\pi^+$ track underwent 
nuclear interaction or decayed in flight
before it came to rest in the RS was estimated 
by comparing Monte Carlo simulations 
with these effects turned on and off. 
The $\pi^+$ acceptance factors of the online and offline cuts
selecting the $\pi^+\to\mu^+\nu$ decay at rest in the RS stopping counter 
and of the $dE/dx$ cuts were measured from 
the sample of scattered beam pions that satisfied the fiducial cuts.
The acceptance factor of the target energy cuts
was measured from the $\pi^+$ tracks in the sample of 
\kpitwo decays tagged by the conversion of both photons from the $\pi^0\to\gamma\gamma$ decay
in the BL calorimeter,
 in order to avoid contamination from shower energy due to 
 photon conversion in the target.
The acceptance factors of the online and offline delayed-coincidence cuts
and the cuts on the \v{C}erenkov counter,
proportional chambers, energy-loss counter and target
were measured from the sample of \kmutwo decays; 
since these cuts were not related to the kinematic values of $\pi^+$,
these factors were assumed to be the same for both the 
\kpigamma and \kpitwo decays.

The acceptance factors 
of the $\gamma$ reconstruction and fiducial cuts 
were estimated from Monte Carlo simulation. 
The acceptance loss of the \kpitwo events due to the 
trigger requirements on the shower and visible energy 
in the BL, endcap and RS counters
(``online photon veto cuts'')
was measured from the sample of \kpitwo decays.
The acceptance loss of \kpigamma
due to the online and offline photon veto cuts
and the RS preshower cut, 
which could detect
a part of the shower from the 227-MeV photon in the \kpigamma decay, 
must be taken into account.
Since a data sample of 227-MeV photons from kaon decays at rest 
was not available,
the acceptance factor was estimated
from studies that compared the performance of these cuts 
on the photons in Monte Carlo simulation 
and on the sample of \kpitwo decays.
The acceptance loss by the accidental hits in the detector subsystems
was measured 
from the sample of \kmutwo decays.
Acceptance losses due to the cuts on the BL hits in the cluster
(the coincidence cut between the times of the photon and the $\pi^+$ track
 and the BL cluster cut)
were confirmed to be negligible 
from the photons in the sample of \kpitwo decays.

With the total acceptances for \kpigamma ($0.0143$) 
and for \kpipi ($0.00357$) in Table~\ref{tab:acc_pgamma},
the number of surviving \kpitwo events ($3.62\times 10^{5}$)
and the \kpitwo branching ratio ($0.2116$), 
the single-event sensitivity for \kpigamma was 
${(1.46\pm 0.09)\times 10^{-7}}$, 
which is four times better than the sensitivity achieved in \cite{Asano3}.
The main source of the error (6\%) was 
due to the systematic uncertainty 
in the acceptance loss of the 227-MeV photon from \kpigamma 
due to the online and offline photon veto cuts
and the RS preshower cut.

In order to verify that the sensitivity for \kpigamma obtained 
from the ratio to \kpitwo decay was correct,  
a branching ratio for \kpitwo 
relative to the sample of \kmutwo decays
was measured 
and the result deviated by $+6$\% from 
the known ratio of the \kpitwo and \kmutwo branching ratios.
We have conservatively
assigned the shift to be an additional systematic uncertainty
in the acceptance of the $\pi^+$ track, 
most probably due to the uncertainty in $\pi^+$ nuclear interaction. 
The total
estimated systematic uncertainty in this search is therefore $8.5$\%.

\section{Conclusion}

A search for the decay \kpigamma was performed 
with the E787 detector at BNL
as a test of angular momentum conservation
in particle physics.
Since no events were observed in the signal region,
in the absence of background
and taking 2.44 events instead of zero
according to the unified approach~\cite{PDB,FeldmanCousins},
we set a 90\% C.L. upper limit $3.6\times 10^{-7}$
on the branching ratio for \kpigamma decay.
The systematic uncertainty was not taken into consideration
in deriving the limit.

The current search is statistically limited, 
and there are good prospects to improve the sensitivity further. 
A new experiment E949~\cite{E949},
which will continue the study of the \pnn decay at BNL,
can yield further gains
by virtue of a larger kaon exposure,
trigger optimization and
improved photon detection capability.

\section*{Acknowledgments}

We gratefully acknowledge the dedicated effort of the technical
staff supporting this experiment and of the Brookhaven 
Collider-Accelerator Department.  
This research was supported in part by the
U.S. Department of Energy under Contracts No. DE-AC02-98CH10886,
W-7405-ENG-36, and grant DE-FG02-91ER40671, by the Ministry of
Education, Culture, Sports, Science and Technology of Japan
through the Japan-U.S. Cooperative Research Program
in High Energy Physics and under the Grant-in-Aids for
Scientific Research, for Encouragement of Young Scientists and for
JSPS Fellows,
and by the Natural
Sciences and Engineering Research Council and the National Research
Council of Canada.

%-----------------------------------------------------------------------------
\pagebreak

%%%%%%%%%%%%%%%%%%%%%%%%%%%%%%begin references%%%%%%%%%%%%%%%%%%%%%%%%%%%%%%%%%

%%%%%%%%%%%%%%%%%%%%%%%%%%%%%%end references%%%%%%%%%%%%%%%%%%%%%%%%%%%%%%%%%

%---------------------------------------------------------------------------
\pagebreak

\begin{table}
\begin{center}
\begin{tabular}{lccl} 
 Acceptance Factors             		& $\pi^+\gamma$	& $\pi^+\pi^0$
	& samples\\
\hline
 $\pi^+$ reconstruction cuts			
						&  $0.399$	& $0.400$
	&MC, $K_{\mu 2}$\\ 
 $\pi^+$ fiducial cuts				
						&  $0.830$	& $0.694$
	&MC\\
 $\pi^+$ stop without nuclear interaction                    
						&  $0.477$	& $0.586$
	&MC\\
 \ \ or decay-in-flight
						&   		&
	&\\
 Transient digitizer ($\pi^+\to\mu^+\nu$) cuts
						&  $0.553$	& $0.592$
	&$\pi_{scat}$\\
 $dE/dx$ cuts			
						&  $0.878$	& ni(*)
	&$\pi_{scat}$\\
 Target energy   cuts			
						&  $0.924$	& ni(*)
	&$K_{\pi 2}$\\
 Other cuts on beam and target 		
						&  $0.606$	& $0.606$
	&$K_{\mu 2}$\\
 $\gamma$ reconstruction and fiducial cuts
						&  $0.693$	& $0.232$
	&MC\\
 Online photon veto cuts to $\pi^+\pi^0$		
                               			&  ni(*)	& $0.264$
	&$K_{\pi 2}$\\
 Online and offline photon veto cuts                   
						&  $0.482$	& ni(*)
	&MC, $K_{\pi 2}$,\\
 \ \ and RS preshower cut	
						&           	&   
	&$K_{\mu 2}$\\
 \ \ to the photon from $\pi^+\gamma$		
						&           	&   
	&\\
\hline
Total acceptance				&  $0.0143$  	& $0.00357$
 	&\\
\end{tabular}
\end{center}
\caption{Acceptance factors for \kpigamma and \kpipi ,
         and the samples used to determine them. 
         ``MC'', ``$K_{\pi 2}$'', ``$K_{\mu 2}$'' and ``$\pi_{scat}$'' mean
	 the sample generated by Monte Carlo simulation and 
         the data samples of \kpipi decays, \kmutwo decays 
	 and scattered beam pions
	 accumulated by calibration triggers, 
         respectively.
         The $dE/dx$ cuts, target energy cuts, offline photon veto cuts
         and RS preshower cut 
         are not imposed on \kpipi .
         (*) not imposed.}
\label{tab:acc_pgamma}
\end{table}

\begin{figure}
\hfill
\centerline{
\hbox{\psfig{figure=e787kpgrev_fig3abw_e.epsi,width=6.5cm}}
\hbox{\psfig{figure=e787kpgrev_fig3b_re.epsi,width=6.5cm}}
}
\hfill
\caption{
  (a)    Kinetic energy distribution of the $\pi\pi^0$ sample.
         The unhatched and hatched histograms represent the distributions
         before and after the photon veto cuts are imposed, respectively.
         The region between the arrows indicates the \kpitwo region .
  (b)     Range versus kinetic energy plot of 
          the events in the subset of the $\pi\gamma$ sample
          tagged by the inverted photon veto cuts.
          The box indicates the \kpigamma signal region.
   $dE/dx$ cuts were imposed on both the 
   $\pi\pi^0$ sample in (a) and the $\pi\gamma$ sample in (b) in advance.}
\label{fig:summary_kpi2}
\end{figure}

\begin{figure}
\hfil
\centerline{
\hbox{\psfig{figure=e787kpgrev_fig4abw_e.epsi,width=6.5cm}}
\hbox{\psfig{figure=e787kpgrev_fig4b_re.epsi,width=6.5cm}}
}
\hfill
\caption{
  (a)    Kinetic energy distribution of the $\pi\pi^0$ sample.
         The unhatched and hatched histograms represent the distributions
         before and after the $dE/dx$ cuts are imposed, respectively.
         The region between the arrows indicates the signal region for \kpigamma .
  (b)     Range versus kinetic energy plot 
          of the events in the subset of the $\pi\gamma$ sample
          tagged by the inverted $dE/dx$ cuts.
          The box indicates the \kpigamma signal region.
   Photon veto cuts were imposed on both the 
   $\pi\pi^0$ sample in (a) and the $\pi\gamma$ sample in (b) in advance.}
\label{fig:summary_overlap}
\end{figure}

\end{document}